\documentclass[conference,a4paper]{IEEEtran}
\IEEEoverridecommandlockouts
\usepackage{cite}
\usepackage{amsmath,amssymb,amsfonts}
\usepackage{graphicx}
\usepackage{epstopdf}
\usepackage{textcomp}
\usepackage{subcaption}
\usepackage{cuted}
\usepackage{hyperref}
\usepackage{multirow}
\usepackage{amsmath}
\usepackage[english]{babel}
\usepackage{amsthm}
\usepackage{comment}
\usepackage{algorithm}
\usepackage{algpseudocode}
\usepackage[table]{xcolor}

\usepackage[table]{xcolor}
\usepackage[left=1.65cm,right=1.65cm,top=1.93cm,bottom=4.5cm]{geometry} 
\makeatletter
\def\ps@IEEEtitlepagestyle{%
  \def\@oddfoot{\mycopyrightnotice}%
}
\def\mycopyrightnotice{%
  \begin{minipage}{\textwidth}
  \centering \scriptsize
  \copyright 2025 IEEE. Personal use of this material is permitted. Permission from IEEE must be obtained for all other uses, in any current or future media, including reprinting/republishing this material for advertising or promotional purposes, creating new collective works, for resale or redistribution to servers or lists, or reuse of any copyrighted component of this work in other works.
  \end{minipage}
}
\makeatother

\def\BibTeX{{\rm B\kern-.05em{\sc i\kern-.025em b}\kern-.08em
    T\kern-.1667em\lower.7ex\hbox{E}\kern-.125emX}}
\begin{document}
\title{Empirical Line-of-Sight Probability Modeling for UAVs in Random Urban Layouts\\
}

\author{Abdul Saboor\textsuperscript{1},
        Zhuangzhuang Cui\textsuperscript{1},
        Evgenii Vinogradov\textsuperscript{1, 2},
        Sofie Pollin\textsuperscript{1}
\\\textsuperscript{1}WaveCoRE of the Department of Electrical Engineering (ESAT), KU Leuven, Leuven, Belgium
\\\textsuperscript{2}Autonomous Robotics Research Center, Technology Innovation Institute, Abu Dhabi, UAE
\\Email:\{abdul.saboor, zhuangzhuang.cui, sofie.pollin\}@kuleuven.be, evgenii.vinogradov@tii.ae}

\maketitle

\begin{abstract}
Accurate Probability of Line-of-Sight ($P_{LoS}$) modeling is important in evaluating the performance of Unmanned Aerial Vehicle (UAV)--based communication systems in urban environments, where real-time communication and low latency are often major requirements. Existing $P_{LoS}$ models often rely on simplified Manhattan grid layouts using International Telecommunication Union (ITU)--defined built-up parameters, which may not reflect the randomness of real cities. Therefore, this paper introduces the Urban Line-of-Sight Simulator (ULS) to model $P_{LoS}$ for three random city layouts with varying building sizes and shapes constructed using ITU built-up parameters. Based on the ULS simulated data, we obtained the empirical $P_{LoS}$ for four standard urban environments across three different city layouts. Finally, we analyze how well Manhattan grid-based models replicate $P_{LoS}$ results from random and real-world layouts, providing insights into their applicability for time-critical communication systems in urban IoT networks.

\end{abstract}


\section{Introduction}
The next generation of 6G networks promises higher data rates and Ultra-Reliable Low-Latency Communications (URLLC) for applications including Intelligent Transportation Systems (ITS), Industrial Internet of Things (I-IoTs), Advanced Air Mobility (AAM), and robotics \cite{chataut20246g}, where time-critical operations are increasingly essential. One major challenge in 6G networks is managing the higher frequency bands, such as millimeter wave and terahertz, which offer high data rates but are more vulnerable to obstacles \cite{Tripathi2021}. In urban environments, tall buildings will likely obstruct these frequencies' Line-of-Sight (LoS), resulting in inconsistent and unreliable communication. Unmanned Aerial Vehicles (UAVs)--assisted Aerial Base Stations (ABSs) are envisioned as an enabler for 6G due to their flexible and on-demand communication infrastructure for coverage with high LoS availability \cite{saboor2021elevating}. However, accurately predicting the Probability of LoS ($P_{LoS}$) between the ABS and User Equipment (UE) is a critical requirement to optimize UAV placement, ensure reliable coverage, and maintain time-critical and high-quality communication links \cite{kim2022use}. $P_{LoS}$ modeling is particularly important in 6G, where LoS availability directly impacts performance metrics such as data rates, latency, and overall network reliability.

Urban environments are highly variable, with unique layouts that include different building areas, heights, shapes, and densities. The International Telecommunication Union (ITU) defines a set of built-up parameters to model urban environments \cite{ITU}. As a result, most existing $P_{LoS}$ models are based on the Manhattan grid structure \cite{ITU, hourani2014, ICC, Saboor2023plos, rev1, rev2}. These models assume uniform street widths, shapes, building dimensions, and regular spacing between buildings. However, the layout provided by the ITU follows a simplified Manhattan grid structure, which does not fully depict the complexity of real urban environments. This simplification can lead to significant inaccuracies when predicting $P_{LoS}$, especially in non-grid-like cities. In contrast, the Ray-Tracing (RT)--based $P_{LoS}$ model provides accurate results for a particular environment \cite{song2022air, li2021ray}. However, RT is computationally expensive, time-consuming, and costly, especially when applied to multiple environments \cite{Saboor2023plos}.   

To address these challenges, this paper introduces the Urban LoS Simulator (ULS) for $P_{LoS}$ modeling between ABS and UEs. The ULS simulates three random urban layouts, where each layout generates cities with varying building shapes, areas, heights, and placements, thus capturing the randomness and complexity of real cities. By applying ITU-defined parameters that control building areas, numbers, and heights, we create different layouts, each with a unique topology. We emphasize that this study focuses specifically on $P_{LoS}$ modeling in ITU-defined environments using built-up parameters, as outlined in \cite{ITU}. Therefore, $P_{LoS}$ models based on stochastic geometry \cite{gapeyenko2021line, al2020probability} are beyond the scope of this work. The key contributions of this paper are:  


\begin{itemize}
    \item We develop ULS that can generate three different layouts for random urban environments using ITU-defined built-up parameters. It helps in more accurate $P_{LoS}$ estimation by selecting the appropriate layout based on the underlying city structure.  
    \item We derive an empirical formula for each layout for predicting $P_{LoS}$ as a function of the elevation angle $\theta$ between the ABS and UE.
    \item We compare the proposed empirical models with existing Manhattan-based models to examine how well the Manhattan grid-based models can reproduce the results obtained from random and real-world city layouts. Furthermore, we conducted a case study using Wireless Insite (WI) to analyze which layout most closely represents $P_{LoS}$ for real urban cities. 
   
\end{itemize}

This paper is structured as follows: Section II reviews the ITU-based Manhattan layout, while Section III presents the proposed three ULS layouts. Section IV explains ULS working and empirical $P_{LoS}$ derivation. Section V discusses the results, and Section VI concludes the paper.

\section{ITU-based Manhattan Urban Layout}
In urban environments, density, height, placement, and area of buildings influence signals' propagation characteristics, particularly $P_{LoS}$. ITU introduced a set of built-up parameters ($\alpha, \beta, \gamma$) to control these characteristics. $\alpha$ represents the total area occupied by buildings in km$^2$, $\beta$ is the number of buildings per km$^2$, and $\gamma$ is the Rayleigh scale parameter to manage buildings' height. Table \ref{tab1} lists the built-up parameters for four standard urban environments.  

\begin{table}[!t]
\caption{Built-up parameters for standard environments.}
\centering
\begin{tabular}
{|l|c|c|c|}
\hline 
\textbf{Environment} & \textbf{$\alpha$} & \textbf{$\beta$ (buildings/km$^2$)}  & \textbf{$\gamma$ (m)} \\ \hline
Suburban & 0.1 & 750 & 8 \\ \hline
Urban & 0.3 & 500 & 15 \\ \hline
Dense Urban & 0.5 & 300 & 20 \\ \hline
Urban High-rise & 0.5 & 300 & 50 \\ \hline
\end{tabular}
\label{tab1}

\end{table}

ITU and existing $P_{LoS}$ models use the Manhattan grid layout, which is a simplified version of a real-world environment \cite{ITU, hourani2014, ICC, Saboor2023plos, rev1, rev2}. The layout assumes all the buildings in the city are square, have the same Width ($W$), and are placed at fixed intervals, as shown in \cite{ITU}. The uniform space between the buildings, or structure-free places, represents Streets ($S$). Given the built-up parameters, the $W$ and $S$ in an urban environment is calculated using the following equation:    
\begin{equation}
\label{eq1}
    W = 1000  \sqrt{\frac{\alpha}{\beta}},  S = \frac{1000}{\sqrt{\beta}} - W.
\end{equation}
Similarly, the height distribution of buildings $f(h)$ can be estimated using \eqref{eq3}.   
\begin{equation}
\label{eq3}
   f(h) = \frac{h}{\gamma^2}e^{-\frac{h^2}{2\gamma^2}}.
\end{equation}
Manhattan-style layout is easy to model and simulate due to its regularity, making it ideal for initial studies of communication systems in urban areas. Therefore, a vast amount of $P_{LoS}$ models in literature use this layout \cite{ITU, hourani2014, ICC, Saboor2023plos, rev1, rev2}. However, this layout oversimplifies real urban environments that often consist of irregular building shapes, dimensions, and streets that do not follow a grid pattern. Such irregularity significantly affects LoS availability and signal propagation, highlighting the need for more realistic modeling approaches to estimate $P_{LoS}$ accurately. Therefore, we present ULS in the following section with the ability to compute $P_{LoS}$ in random urban layouts.      

\section{Proposed Layouts in ULS}
To address the limitations of the abovementioned Manhattan grid structure, we propose three new urban layouts of random urban environments with varying building sizes, shapes, heights, and distributions. The following sections will explain each layout. 


\begin{figure*}[!t]
  \centering
    \begin{subfigure}[b]{0.3\linewidth}
    \includegraphics[width=\linewidth]{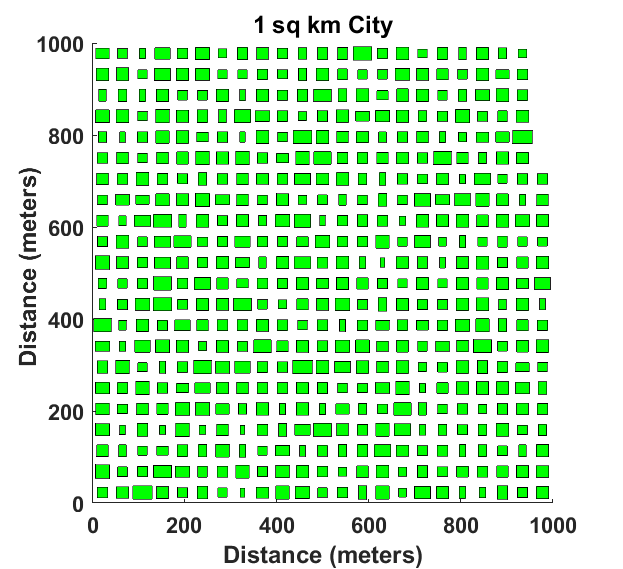}
    \caption{Random-Manhattan Layout}
    \label{sim1}
  \end{subfigure}
  \centering
  \begin{subfigure}[b]{0.3\linewidth}
    \includegraphics[width=\linewidth]{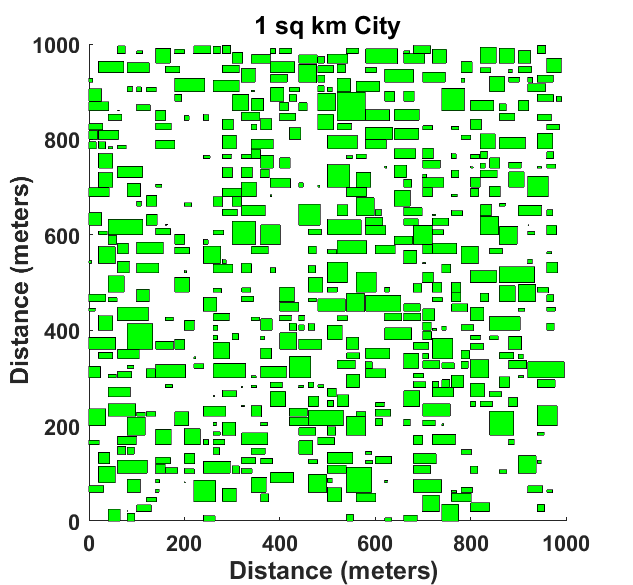}
    \caption{Random-Urban Layout}
    \label{sim2}
  \end{subfigure}
  \centering
  \begin{subfigure}[b]{0.28\linewidth}
    \includegraphics[width=\linewidth]{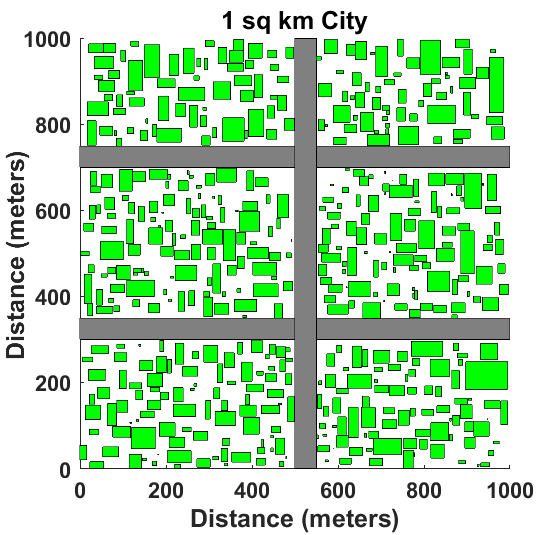}
    \caption{Random-Highway Layout}
    \label{sim3}
  \end{subfigure}
  \caption{Top view of the proposed layouts in ULS for the standard urban environment ($\alpha = 0.3, \beta = 500, \gamma = 15$). }
  \label{fig4}
  \vspace{-1em}
\end{figure*}

\subsection{Random-Manhattan Layout}

The first layout in ULS is the Random-Manhattan (RM) layout, a modified version of the Manhattan layout where buildings are placed in a regular grid structure. However, in contrast to fixed building sizes, each building in the RM layout has a randomly generated size and shape, as given in Fig. \ref{sim1}. The motivation for this simulation is that real-world buildings are not uniformly square, nor are the spaces between them always consistent, even within a Manhattan grid structure. Let's assume $A_{\text{total}}$ is the total area (1 km$^2$ in ULS), the area occupied by buildings $A_{\text{building}}$ can be calculated using:

\begin{equation}
\label{alpharea}
    A_{\text{building}} = \alpha \times A_{\text{total}}
\end{equation}

The average area of each building $B_{\text{avg}}$, is derived from  $A_{\text{building}}$ and $\beta$, where $B_{\text{avg}} = \frac{A_{\text{building}}}{\beta}$. For a Manhattan-style layout, the simulator divides the area into a grid with $n_x$ and $n_y$ blocks along the x-axis and y-axis, respectively. Both values are computed as follows: $n_x = \lceil \sqrt{\beta} \rceil, \quad n_y = \left\lceil \frac{\beta}{n_x} \right\rceil$


The next step is to place buildings around the grid. For that, each building's area is selected randomly around the average area to represent the building logically. The area of $ith$-building ($A_i$) can be calculated as follows: 

\begin{equation}
A_i = B_{\text{avg}} \times (0.6 + 0.8 \cdot r)
\end{equation}
where $r$ is a random variable uniformly distributed between 0 and 1 that ensures that $A_i$ can take a random value between $\left[0.6 \times B_{\text{avg}}, 1.4 \times B_{\text{avg}}\right]$. 

The RM layout considers only square and rectangle building shapes, where the width and length of $ith$ square building is $W_i = L_i = \sqrt{A_i}$. Height can be estimated using equation \eqref{eq3}. In contrast, the width of a rectangular building can be determined using random factor $r$, where $W_i = \sqrt{A_i} \times (0.5 + r)$ and $L_i = \frac{A_i}{W_i}$. In the end, each building is placed in the grid at the center of the designated block area, where $W_{\text{block}} = \frac{1000}{n_x}$ and $L_{\text{block}} = \frac{1000}{n_y}$ represents the average width and length of the block containing building area. The coordinates of $ith$ building $(x_i, y_i)$ in the Manhattan grid are:   

\begin{equation}
\begin{aligned}
x_i &= (i_x - 1) \times W_{\text{block}} + \frac{W_{\text{block}} - W_i}{2}, \\
y_i &= (i_y - 1) \times L_{\text{block}} + \frac{L_{\text{block}} - L_i}{2}
\end{aligned}
\end{equation}

where \( i_x \) and \( i_y \) are the indices of the block in the grid.

The proposed RM layout adds a layer of complexity to the uniform Manhattan grid layout by considering variable building sizes and shapes, considering that even grid-based structures rarely have buildings of identical size. This simulator is suitable for evaluating the potential of ABS or communication infrastructure in grid-structured cities with varied building sizes.  

\subsection{Random-Urban Layout}
Unlike a grid structure, the Random-Urban (RU) layout generates totally random building placements with varying sizes in a km$^2$ area, as shown in Fig. \ref{sim2}. The objective of the RU layout is to focus on a more flexible representation of urban structures. Similar to the RM layout, $A_{\text{total}}$ is one km$^2$ in RU layout, and $A_{\text{building}}$ can be computed using equation \eqref{alpharea}. However, it follows the Dirichlet distribution to determine the areas of the individual buildings. Let the vector  $\mathbf{A} = [A_1, A_2, \dots, A_\beta]$ represent the areas of the $\beta$ buildings. The areas for each building are generated by:
\begin{equation}
\mathbf{A} \sim \text{Dirichlet}(\mathbf{1}_\beta) \times A_{\text{building}}
\end{equation}

Here \( \text{Dirichlet}(\mathbf{1}_\beta) \) generates random building areas from a Dirichlet distribution, and these areas are summed together to the $A_{\text{building}}$, such as $\sum_{i=1}^{\beta} A_i = A_{\text{building}}$.

The RU layout uses a grid approach with a grid resolution $G_r$ = 50, dividing the total area into grid cells of size $\Omega$: $\Omega = \frac{1000}{G_r}$. Each $ith$ building is placed into the grid, and its dimensions are converted to the number of cells required to fit it. Remember, Dirichlet distribution generates varying building areas. Thus, each building can have a different length and width. Let the width $W_i$ and length $L_i$ of a building be converted into cells as follows:

\begin{equation}
C_x = \left\lceil \frac{W_i}{\Omega} \right\rceil, \quad C_y = \left\lceil \frac{L_i}{\Omega} \right\rceil
\end{equation}
where $C_x$ and $C_y$ are the number of grid cells required in the $x$ and $y$ directions, respectively. For square buildings $W_i = L_i = \sqrt{A_i}$ and the height follows Rayleigh distribution. In contrast, the width and length of a rectangular building would be $W_i = 1.5 \times \sqrt{A_i}, \quad L_i = \frac{A_i}{W_i}$. Additionally, the simulator ensures that a specific building should not be greater than 3\% of the total building area using $0.03 \times A_{\text{total}}$.   

Each building is randomly placed in the grid, and its placement is constrained to ensure no overlap with existing buildings. Let $G(x, y)$ be the grid matrix, where each entry is `0' if the cell is free and `1' if occupied. The building is randomly placed at $ (x_i, y_i) $ if:

\begin{equation}
\footnotesize
\forall x, y \in \{x_i, \dots, x_i + C_x - 1\}, \{y_i, \dots, y_i + C_y - 1\},  G(x, y) = 0
\end{equation}

If such a position is found, the corresponding grid cells are marked as occupied:

\begin{equation}
G(x, y) = 1 \quad \text{for all occupied cells}
\end{equation}
This RU layout is a better choice for ABS performance evaluations for cities that do not follow a strict grid structure or have many irregularities, such as old European cities or some residential areas.

\subsection{Random-Highway Layout}
The Random-Highway (RH) layout is a modified version of the RU layout that can define highways in a city, as shown in Fig. \ref{sim3}. First, the proposed simulator defines $n$ highways with the width of $ith$ highway $W^H_i$ and length $L^H_i$. After that, random buildings are placed in the remaining free area, following the same principle as the RU layout. The simulator also ensures that buildings do not overlap with the existing buildings or highways. This layout suits random city structures, with a few highways supporting fast-moving vehicles. Furthermore, the RH layout can separately estimate $P_{LoS}$ for moving vehicles on the highway and UEs distributed in streets across the city.

\begin{algorithm}[!t]
\small
\caption{$P_{LoS}$ estimation in ULS.}
\label{algo1}
\begin{algorithmic}
\State \textbf{Input:} $\alpha$, $\beta$, $\gamma$, $h^{max}_{ABS}$, $n_{UE}$, $n_{cities}$
\State \textbf{Output:} $P_{LoS}(\theta)$

\State Initialize $\textit{los\_sum} \gets \mathbf{0}_{1 \times 91}$, $\textit{los\_count} \gets \mathbf{0}_{1 \times 91}$

\For{i = 1 \textbf{to} $n_{cities}$}
    \State [buildings, $A_{\text{total}}$] $\gets$ \text{layout}($\alpha$, $\beta$, $\gamma$) \\
    \text{ // Generate city layout and buildings using any layout}
    \While{true}
        \State [ABS$_x$, ABS$_y$] $\gets$ \text{random coordinates in city} 
        \State $h_{ABS}$ $\gets$ \text{rand} $\times h^{max}_{ABS}$
        \If{\text{check\_collision}}
            \State \textbf{break} \text{ // Exit loop if no collision detected}
        \EndIf
    \EndWhile  
    \State [UE$_x$, UE$_y$] $\gets$ \text{random coordinates in city}
    \\ \text{//Place random $n_{UE}$ in city without building collisions}
    \State \textbf{Compute} $\gets$ horizontal distance and $\theta$ for each UE  
    \State \textbf{Compute} $\gets$ $LoS$ for all UEs
    \For{$\theta$ = 0 \textbf{to} 90}
        \State \text{UE} $\gets$ \text{angles} == $\theta$ \text{ // Identify UEs at this angle}
        \If{\text{any UE of particular $\theta$}}
            \State $los\_sum(\theta)$ $\gets$ Add the calculated $LoS$ value to existing sum for that $\theta$ 
            \State $los\_count(\theta)$ $\gets$ Increment $Count$ for that $\theta$ 
        \EndIf
    \EndFor
\EndFor

\State $P_{LoS}(\theta)$ $\gets$ $los\_sum(\theta)$ $\div$ ${los\_count(\theta)}$ 
\State \textbf{return} $P_{LoS}(\theta)$
\end{algorithmic}
\end{algorithm}

\section{ULS working flow and Empirical $P_{LoS}$}
The complete simulation process of ULS to compute $P_{LoS}(\theta)$ according to elevation angle $\theta$ is given in Algorithm \ref{algo1}. The simulation starts by generating a city using the built-up parameters ($\alpha$, $\beta$, and $\gamma$). The city's layout depends on the underlying module (RM, RU, or RH). After that, it randomly places an ABS at altitude $h_{ABS}$ between 0 and $h^{max}_{ABS}$, ensuring it does not collide with any buildings. The $n$ UE's ($n_{UE}$) are placed on the ground randomly within the city boundaries. The ULS ensures that no UE resides inside the building for a fair $P_{LoS}(\theta)$ averaging. For each UE, the ULS computes the distance $r$ and the corresponding $\theta$ with ABS. Later, it checks whether any building between UE and ABS obstructs the LoS. If there are $k$ buildings between UE and ABS, and $ith$ building is at distance $r_1$, the obstruction height of that building $h_{OB(i)}$ is calculated using the following equation \cite{10299705}:

\begin{equation}
\label{ITU}
    h_{OB(i)} = h_{ABS} - \frac{r_{i}\times (h_{ABS}-h_{UE})}{r}.
\end{equation}

In the above equation, $h_{UE}$ represents UE height. The ULS compares each $h_{OB(i)}$ height with the underlying $ith$ building height $h_{B(i)}$, and the link is only considered LoS if all $h_{OB(i)} > h_{B(i)}$. After that, the ULS updates the $los\_sum$ and ${los\_count}$  for the respective elevation angle. Once the simulation is complete, the average $P_{LoS}(\theta)$ for each angle is computed by dividing the accumulated $los\_sum$ by the number of observations $los\_count$. 

Finally, we use the Sigmoid function based on the simulation data to derive smooth empirical $P_{LoS}(\theta)$ for four standard environments. The Sigmoid function fluctuates between 0 and 1, which makes it well-suited for modeling $P_{LoS}$. In this paper, we use two Sigmoid functions for comparison. The first Sigmoid function ($Sig_1$) is a well-known function presented in \cite{hourani2014}.

\begin{equation} 
\label{CF}
P_{LoS}(\theta^\circ) = \frac{1}{1 + a \cdot \exp(-b \cdot (\theta^\circ - a))} 
\end{equation}

Where $a$ and $b$ are empirically derived coefficients from the simulation data to ensure the best fit to $P_{LoS}(\theta^\circ)$ values across different urban environments and layouts. The second Sigmoid function ($Sig_2$) assumes the $\theta$ in \textbf{radians}, as given in equation \eqref{CF1} 

\begin{equation} 
\label{CF1}
P_{LoS}(\theta) = \frac{1}{1 + \exp( x_1\theta^3 +  x_2\theta^2 + x_3\theta + x_4   )} 
\end{equation}

where, $x_1, x_2, x_3$ and $x_4$ are fitting parameters. The primary goal of these empirical models is to provide a simplified closed-form representation to accurately estimate $P_{LoS}$ for UAV-assisted communication in different urban environments. The following section will compare the fitting performance of both empirical models and provide fitting parameters for the four standard urban environments.

\begin{figure}[!t]
\centerline{\includegraphics[width=.6\linewidth]{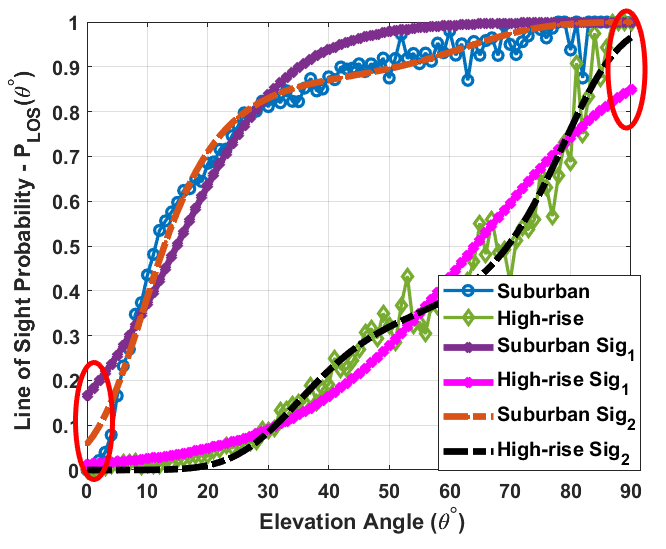}}
\caption{$Sig_1$ and $Sig_2$ fitting comparison.}
\label{sig1vs2}
\vspace{-2em} 
\end{figure}

\begin{figure*}[!t]
  \centering
    \begin{subfigure}[b]{0.25\linewidth}
    \includegraphics[width=\linewidth]{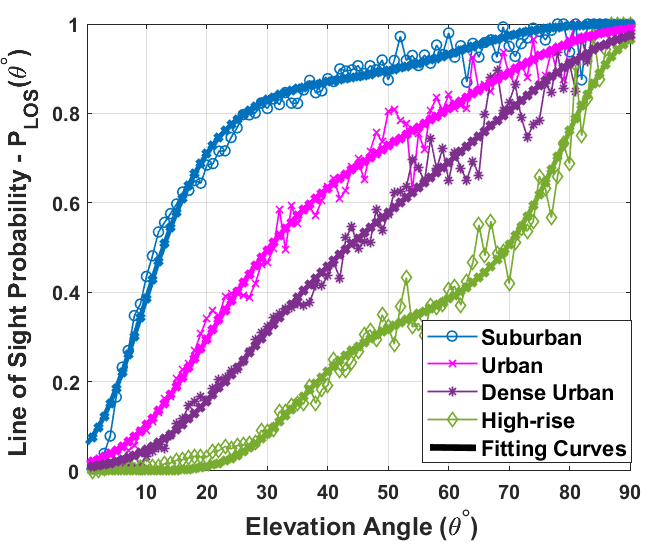}
    \caption{RM layout}
    \label{RM Curves}
  \end{subfigure}
  \centering
  \begin{subfigure}[b]{0.25\linewidth}
    \includegraphics[width=\linewidth]{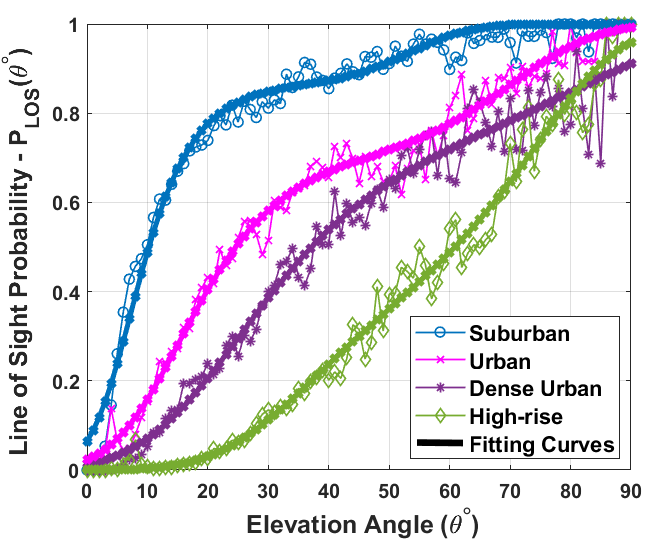}
    \caption{RU layout}
    \label{RU Curves}
  \end{subfigure}
  \centering
  \begin{subfigure}[b]{0.25\linewidth}
    \includegraphics[width=\linewidth]{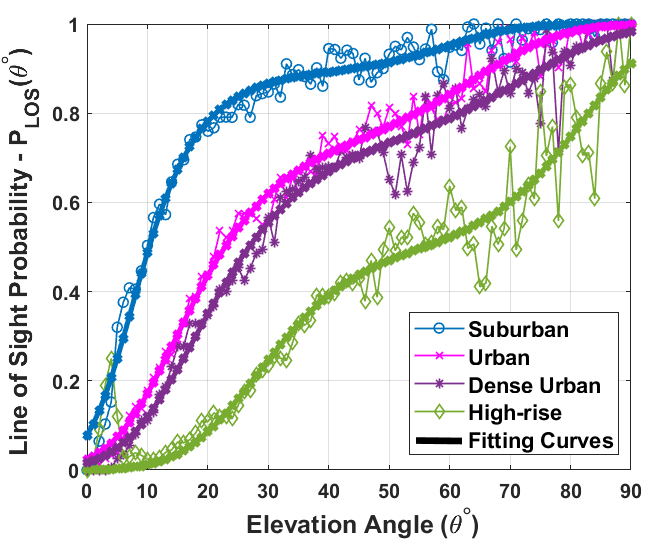}
    \caption{RH layout}
    \label{RH Curves}
  \end{subfigure}
  \caption{Simulated and empirical $P_{LoS}(\theta)$ curves for the four standard urban environments using different layouts.}
  \label{RMUH curves}
\end{figure*}

\section{Results and Analysis}
This section uses the proposed layout modules to compare the $P_{LoS}$ for four standard environments. In each simulation, we put 5K random UEs with height zero inside the city, and 40 random cities are simulated for each urban environment under a specific layout. Therefore, 200K points between ABS-UE are taken for averaging. 

The first objective is determining the most suitable Sigmoid function for $P_{LoS}$ fitting. Therefore, we compare the suburban and high-rise simulated results in RM layout with both $Sig_1$ and $Sig_2$, as shown in Fig \ref{sig1vs2}. \textbf{\emph{While $Sig_1$ generally aligns with the simulated data, it shows notable deviations at lower elevation angles in suburban environments and at higher elevation angles in high-rise settings, as highlighted in the figure.}}  In contrast, the fitting results for $Sig_2$ are better with a Root Mean Square Error (RMSE) of 0.0340 for suburban and 0.0399 for high-rise environments compared to $Sig_1$ with RMSE of 0.0687 for suburban and 0.0664 for high-rise environments. The main reason behind the superior performance of $Sig_2$ is its cubic polynomial form that offers better flexibility in modeling $P_{LoS}$ variations compared to $Sig_1$'s simpler exponential structure. Given $ Sig_2$'s superior performance, we will only use it for empirical fitting in the remaining results to maintain clarity in the figures and avoid unnecessary clutter.      

Fig. \ref{RM Curves} plots $P_{LoS}$ for the four standard urban environments using RM layout. \textbf{Thin} lines in the figure represent the Monte-Carlo simulation results for 200K ABS-UE combinations. In contrast, \textbf{thick} lines are the fitted S-curves derived using MATLAB's curve fitting toolbox based on the equation \eqref{CF1}. The values of fitting parameters for $Sig_1$ and $Sig_2$ are given in Table \ref{tab2}. The results are consistent with the existing studies, where the UEs in suburban environments experience the highest average $P_{LoS}$ and high-rise environment offers the lowest $P_{LoS}$. \textbf{\emph{One main observation is the influence of building heights on $P_{LoS}$ is more prominent compared to the buildings area $A_{\text{building}}$. Therefore, the $P_{LoS}$ difference between urban and dense urban environments is marginal compared to the suburban and high-rise environments.}} The same trend is followed in RU and RH layouts, as shown in Fig. \ref{RU Curves} and Fig. \ref{RH Curves}.

\begin{table}[!t]
\caption{$Sig_1$ and $Sig_2$ S-curve fitting parameters for the four standard urban environments using different layouts.}
\centering
\begin{tabular}
{|l|c|c|c|c|c|c|}

\hline 
\multirow{3}{*}{\textbf{Environment}} & \multicolumn{6}{c|}{\textbf{RM Layout}}  \\ \cline{2-7}
 & \multicolumn{2}{c|}{\textbf{$Sig_1$}} & \multicolumn{4}{c|}{\textbf{$Sig_2$}} \\ \cline{2-7}
& \textbf{$a$} & \textbf{$b$} & \textbf{$x_1$} & \textbf{$x_2$} & \textbf{$x_3$} & \textbf{$x_4$} \\ \hline 
Suburban & 3.44 & 0.108 & -9.31  & 20.71  & -16.64  & 2.78  \\ \hline
Urban & 6.55 & 0.069 & -4.933  & 12.4  & -12.83  & 4.049  \\ \hline
Dense Urban & 9.67 & 0.064 & -4.253  & 11.13  &  -12.37  & 4.827  \\ \hline
High-rise & 19.8 & 0.067 & -13.16  & 37.89  & -37.91  & 13.73 \\ \hline
\multicolumn{7}{|c|}{\textbf{RU Layout}}  \\ \hline
Suburban  & 2.96 & 0.117 &  -16.54 & 30.55  & -19.85  & 2.668 \\ \hline
Urban  & 4.42 & 0.056 & -6.686  &  16.24  & -14.42  & 3.726 \\ \hline
Dense Urban  & 7.06 & 0.056 & -2.772  & 8.748 & -11.10  & 4.276 \\ \hline
High-rise  & 18.4 & 0.071 & -6.721  & 18.93  & -20.69  & 8.675 \\ \hline
\multicolumn{7}{|c|}{\textbf{RH Layout}}  \\ \hline
Suburban &  2.99 & 0.124 & -11.49 & 23.93  & -17.67  &   2.468 \\ \hline
Urban & 4.72 & 0.068 & -7.536  & 17.33  & -15.02  & 3.709 \\ \hline
Dense Urban &  5.30 & 0.063 & -5.589  & 14.63  & -14.35  & 4.083 \\ \hline
High-rise  & 9.24 & 0.048 & -7.308  & 21.05  & -21.34 &  7.568  \\ \hline
\end{tabular}
\vspace{-1.5em}
\label{tab2}

\end{table}

\begin{figure}[!t]
    \centering
    \begin{minipage}{0.49\linewidth}
        \centering
        \includegraphics[width=\linewidth]{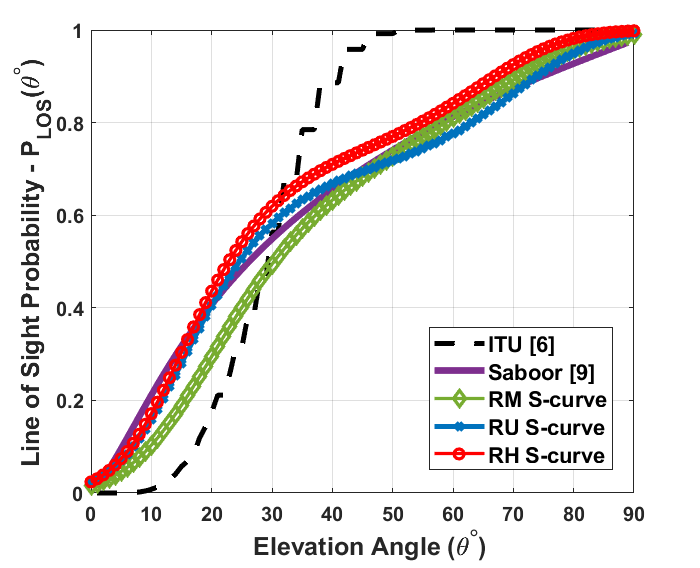}
        \caption{$P_{LoS}$ comparison of Manhattan-based models and simulated layouts in the urban environment.}
        \label{urbanC}
    \end{minipage}
    \hfill
   \begin{minipage}{0.49\linewidth}
        \centering
        \includegraphics[width=\linewidth]{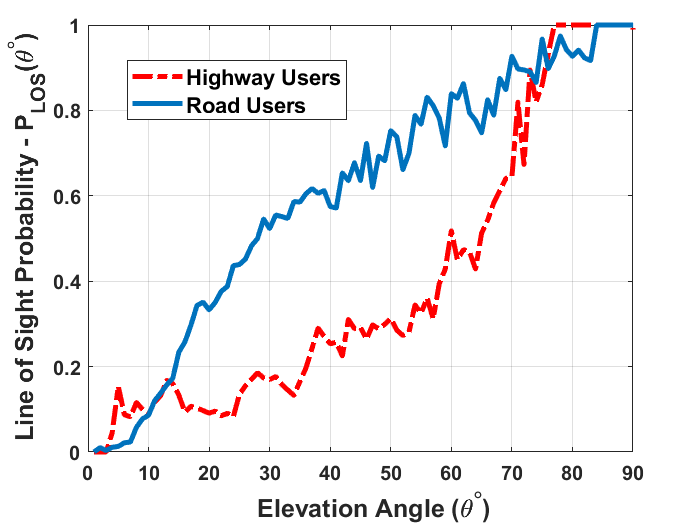}
        \caption{$P_{LoS}$ Comparison for road and street users in RH-based urban environment.}
        \label{roadsvsstreet}
    \end{minipage}
    \vspace{-2em}
\end{figure}

Fig. \ref{urbanC} compares the $P_{LoS}$ of Manhattan-based models \cite{ITU, Saboor2023plos} and simulated layouts in the urban environment. The key observation is that $P_{LoS}$ of all the proposed layouts shows similar trends with the Manhattan-based $P_{LoS}$ model presented in \cite{Saboor2023plos}. \textbf{\emph{This suggests that different layouts do not significantly influence $P_{LoS}$ when constructed with the same built-up parameters.}} Therefore, Manhattan-based $P_{LoS}$, such as \cite{Saboor2023plos}, can provide reasonably accurate results for different random urban layouts and vice versa. 

In Fig. \ref{roadsvsstreet}, we analyze the $P_{LoS}$ for users on highways and streets using RH layout. For analysis, we consider only three highways with 50m width and 1000m length, as illustrated in Fig. \ref{sim3}. \textbf{\emph{The key observation is that with a random deployment of ABS, users on the streets experience better $P_{LoS}$ compared to the users on highway. However, highway users experience very high and stable $P_{LoS}$ at higher angles due to fewer chances of obstruction from buildings, concluding that ABSs should not be deployed far away from highways to support vehicular users on highways requiring more stable channels}}. Furthermore, road users also show better $P_{LoS}$ at smaller angles, mainly because of certain ABS-UE combinations, which reside on highways and thus experience 100\% LoS. However, the results of these findings are only limited to a layout with three fixed highways of the same dimensions. Therefore, more analysis is required using multiple highways at different locations.       

\begin{figure}[!t]
  \centering
  \includegraphics[width=.7\linewidth]{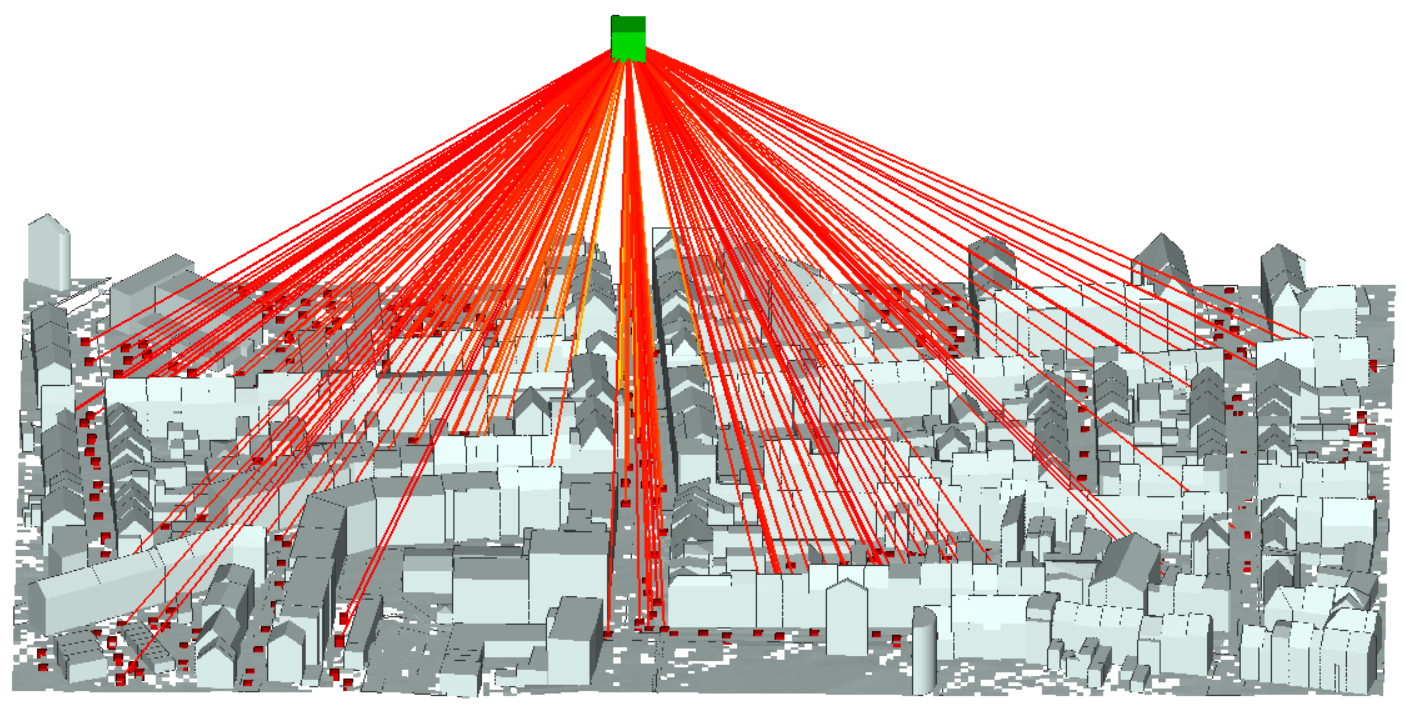}
  \caption{Wireless InSite setup.}
  \label{WI}
  \vspace{-1em}
\end{figure}

\begin{figure}[!t]
  \centering
  \includegraphics[width=.65\linewidth]{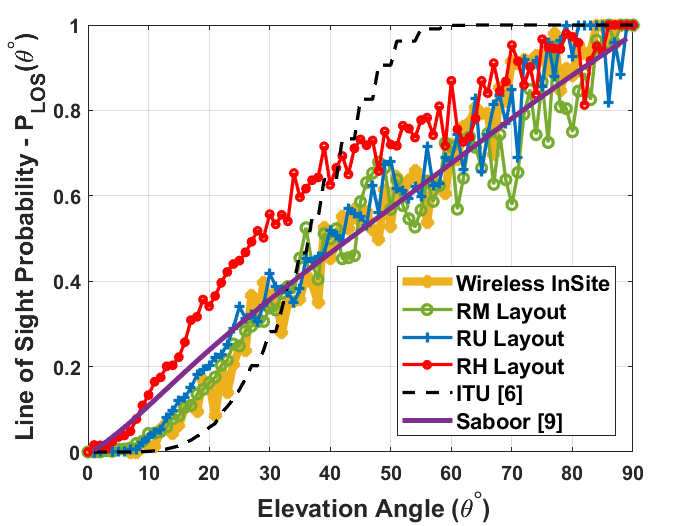}
  \caption{Comparison of proposed urban layouts and Manhattan-based models with Cologne RT data.}
  \label{CologneSim}
\end{figure}

 Finally, we use a ray-tracer WI to model a part of the Cologne residential area. The overall setup is illustrated in Fig. \ref{WI}, where we follow the same approach of ULS by placing an ABS at random height and users at random locations. In total, we get more than 10K ABS-UE combinations to compute $P_{LoS}$. At the same time, we gather the built-up parameters for the observed area ($\alpha$ = 0.55, $\beta$ = 680, $\gamma$ = 12.69) to simulate it in different layouts using ULS. Fig. \ref{CologneSim} compares the proposed urban layouts and Manhattan-based models with Cologne WI data. Table \ref{cologneTab} shows that RU layout provides the best match, with the lowest RMSE of 0.0699, Mean Absolute Error (MAE) of 0.0540, and the highest R$^2$ (0.952). This indicates that the RU layout closely replicates the real-world LoS behavior in residential areas. The Manhattan-based $P_{LoS}$ model \cite{Saboor2023plos} follows, with slightly higher RMSE and MAE, while the RM layout is in third place. In contrast, the ITU-based model and RH have higher error metrics, likely due to the simplicity and lack of highway-specific considerations in the real city model.

\begin{table}[!t]
    \centering
    \caption{RT-based Cologne $P_{LoS}$ comparison with existing models and different layouts. }
    \begin{tabular}{|c|c|c|c|}
        \hline
        \textbf{Model} & \textbf{RMSE} & \textbf{MAE}  & \textbf{R$^2$} \\
        \hline
        Cologne City & ------- & ------- & ------- \\ \hline
        ITU \cite{ITU} & 0.1857  & 0.1427 & 0.658 \\ \hline
        Saboor \cite{Saboor2023plos} & 0.0709 & 0.0610 & 0.949 \\ \hline
        RM Layout & 0.0887 & 0.0657 & 0.922 \\ \hline
        RU Layout & \cellcolor{yellow!30}0.0699 & \cellcolor{yellow!30}0.0540 & \cellcolor{yellow!30}0.952 \\ \hline
        RH Layout & 0.1519 & 0.1289 & 0.772 \\ \hline
    \end{tabular}
    \label{cologneTab}
\end{table}

\section{Conclusion}
This paper introduced the ULS to model $P_{LoS}$ for UAV-assisted communication systems within urban environments. The ULS considers three random urban layouts incorporating varying building sizes, shapes, heights, and distributions for more realistic $P_{LoS}$ modeling. Through extensive simulations, the paper derived empirical $P_{LoS}$ models for four standard urban environments in each layout. The results show that the impact of building height on $P_{LoS}$ is more significant than building area. Furthermore, $P_{LoS}$ for a specific urban layout differs slightly from other layouts, highlighting that ITU-based Manhattan layout-specific models can be used for $P_{LoS}$ analysis in any random layout and environment given the same built-up parameters. Lastly, the WI results show that the RU layout most accurately reflected real-world $P_{LoS}$ behavior for residential areas. The ULS will soon be available for research and contribution to the wireless communication domain.

\section*{Acknowledgment}
This research is supported by the Research Foundation
Flanders (FWO), project no. G0C0623N, and iSEE-6G project
under the Horizon Europe Research and Innovation program
with Grant Agreement No. 101139291.

\bibliographystyle{IEEEtran}
\bibliography{ref}

\end{document}